\documentclass[twocolumn,showpacs]{revtex4}

\usepackage{graphicx}
\usepackage{dcolumn}
\usepackage{amsmath}
\begin{document}

\title{Dirac nodal pockets in the antiferromagnetic parent phase of FeAs superconductors 
}

\author{N.~Harrison$^1$ \& S.~E.~Sebastian$^2$ 
}
\affiliation{
$^1$Los Alamos National Laboratory, MS-E536, Los Alamos, NM 87545\\
$^2$Cavendish Laboratory, Cambridge University, JJ Thomson Avenue, Cambridge CB3~OHE, U.K
}
\date{\today}

\begin{abstract}
We show that previously measured small Fermi surface pockets within the antiferromagnetic phase of SrFe$_2$As$_2$ and BaFe$_2$As$_2$ are consistent with a Dirac dispersion modulated by interlayer hopping, giving rise to a Dirac point in $k$-space and a cusp in the magnetic field angle-dependent magnetic quantum oscillation frequencies. These findings support the existence of a nodal spin-density wave in these materials, which could play an important role in protecting the metallic state against localization effects. %We show that further angle-dependent measurements of the cyclotron effective mass and quantum oscillation amplitude can potentially be used to explore the properties of the quasiparticles close to the Dirac point. 
The speed of the Dirac fermions in SrFe$_2$As$_2$ and BaFe$_2$As$_2$ is found to be 14$-$20 times slower than in graphene, suggesting that the pnictides provide a laboratory for exploring the effects of strongly interacting Dirac fermions.
\end{abstract}
 \maketitle

\begin{figure}
\centering
\includegraphics*[width=0.48\textwidth,angle=0]{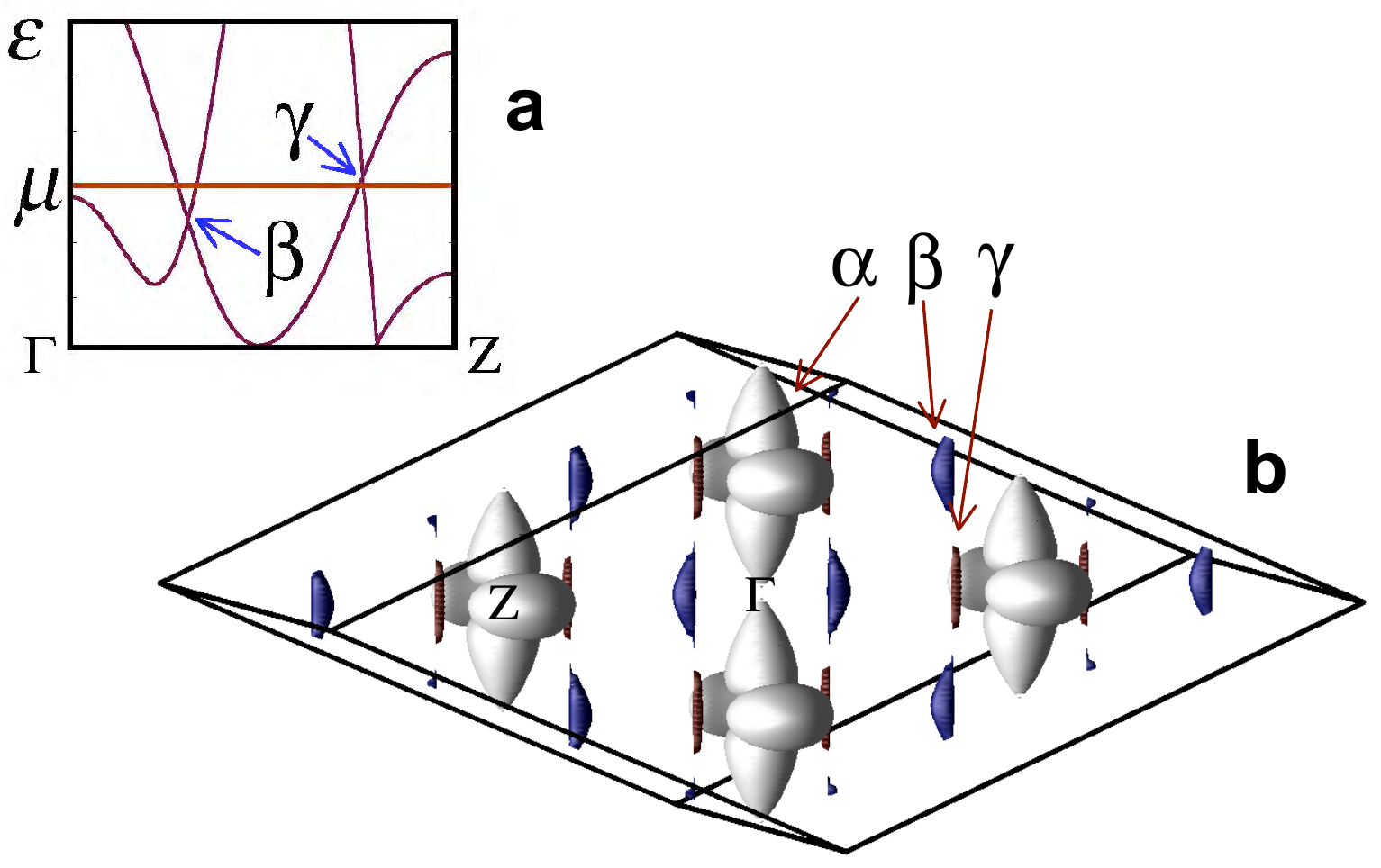}
\caption{Portion of the bandstructure of antiferromagnetic BaFe$_2$As$_2$  ({\bf a}) indicating Dirac points (shifted relative to $\mu=0$) responsible for $\beta$ and $\gamma$ pockets shown in {\bf b}. Calculation provided courtesy of M.~Johannes using the procedure detailed in Ref.~\cite{analytis1}.
}
\label{fig1}
\end{figure}

Despite broad similarities in their superconducting phase diagrams~\cite{kivelson1}, a crucial difference lies in the metallic character of the parent antiferromagnetic phase of the iron arsenide (FeAs) superconductors unlike the cuprates~\cite{sebastian1,analytis1,chen1}. An unresolved question in these materials concerns whether the absence of a Mott insulating regime is the product of greatly reduced electron correlations compared to the cuprates~\cite{sebastian1,si1}. An alternate possibility is a nodal dispersion that protects a metallic density of states at the Fermi energy against complete gapping arising from a strongly correlated spin-density wave~\cite{ran1}. Recent angle-resolved photoemission spectroscopy (ARPES) measurements~\cite{richard1} reveal a conical Dirac-like dispersion within the antiferromagnetic phase of BaFe$_2$As$_2$ (see Fig.~\ref{fig1}a), presenting a potential route for realisation of the latter case. Such a Dirac-like dispersion signals degeneracy in the electronic structure$-$ like graphene~\cite{katsnelson1}, suggesting potential topological quantisation.

While in two-dimensional (2D) graphene a precise tuning of the chemical potential ($\mu$) is required in order to access the Dirac node, hopping along the interlayer direction in the quasi-2D FeAs materials provides an intrinsic tuning to access pockets of fermions positioned at the Dirac nodes. Degenerate electron and hole pockets are expected at the Dirac nodes in such quasi-2D systems, evincing characteristic topology. In this paper, we revisit the small $\beta$ and $\gamma$ Fermi surface pockets observed via quantum oscillations in the antiferromagnetic phase of the FeAs superconductors (SrFe$_2$As$_2$ and BaFe$_2$As$_2$)~\cite{sebastian1,analytis1}  and investigate their origin. In contrast to graphene where Dirac points arise within the original band structure, we look for potential association of the small Fermi surface pockets with Dirac points occurring as a consequence of spin-density wave (SDW) band reconstruction in the FeAs materials~\cite{ran1}. We probe for consistency of the measured $\beta$ and $\gamma$ pockets (bandstructure calculation shown in Fig.~\ref{fig1}) with almost degenerate electron and hole pockets located at a Dirac node. By a comparison of experimentally measured Fermi surface topology and effective mass with predictions, we show that the small $\beta$ and $\gamma$ pockets are consistent with a Dirac dispersion with characteristic speed $c^\ast\sim$~5$-$7~$\times$~10$^4$~ms$^{-1}$ matching that reported in ARPES experiments~\cite{richard1}. 

\begin{figure}
\centering
\includegraphics*[width=0.48\textwidth,angle=0]{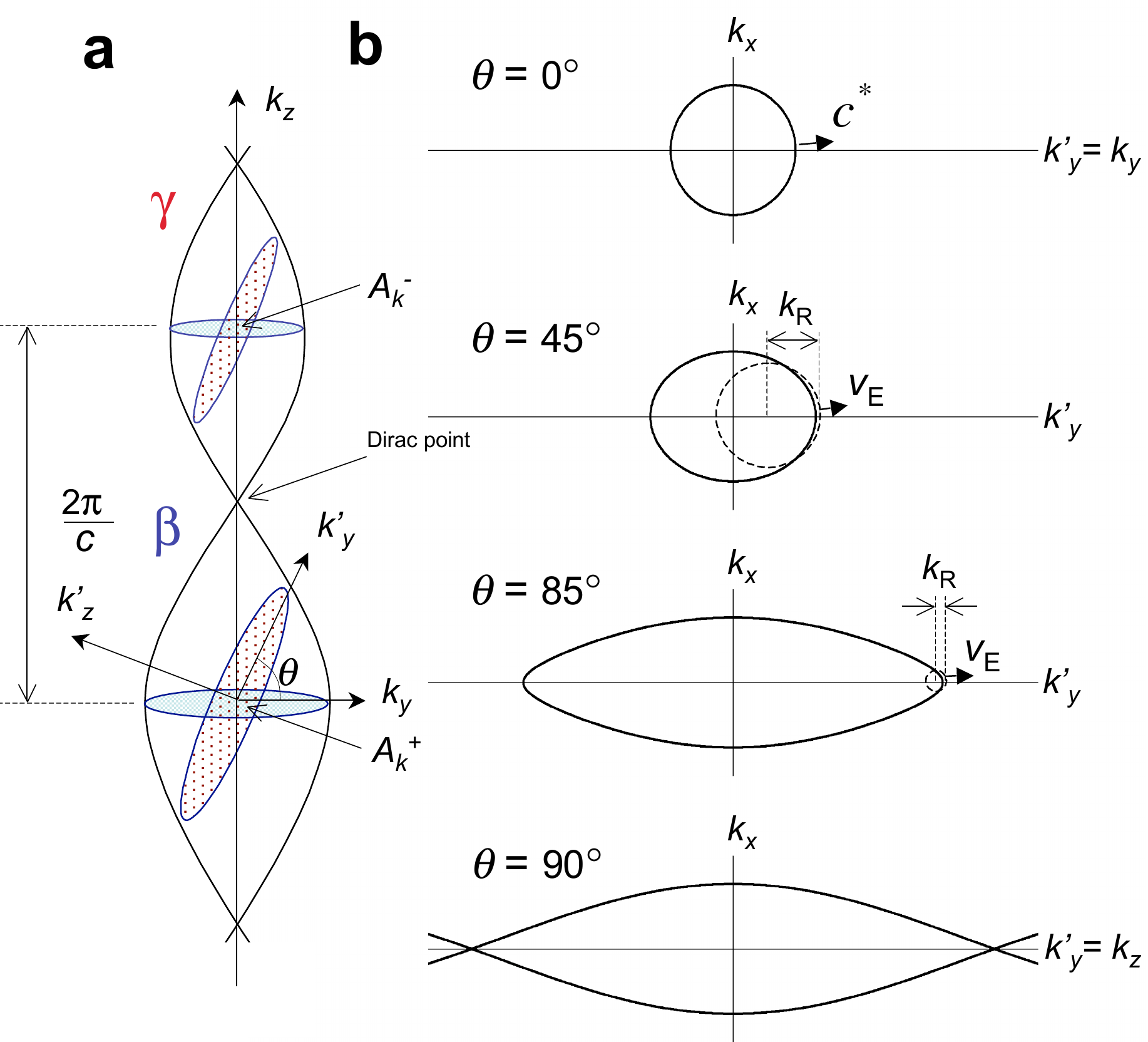}
\caption{{\bf a}, Fermi surface schematic according to the dispersion given by Eqn. \ref{dirac}, showing electron and hole pockets with the planes of the extremal orbits shown for an arbitrary angle $\theta$ between $k_y$ and $k^\prime_y$. The magnetic induction (not shown) is orthogonal to the planes of the orbits. {\bf b}, The shape of the orbit for several different angles $\theta$, indicating the collapse in the radius of curvature $k_{\rm R}$ at its furthest extremity approaching the Dirac point as $\theta\rightarrow$~90$^\circ$.
}
\label{FS}
\end{figure}

Figure~\ref{FS}a depicts how a small interlayer hopping of $t\sim$~10~meV would cause the apex of the Dirac dispersion to intersect the Fermi energy even for a small chemical potential, giving rise to electron and hole pockets as a function of the interlayer dispersion. Vertically stacked electron and hole pockets of the form depicted in Fig.~\ref{FS}a touch at their extrema, representing the Dirac point in $k$-space. By considering a minimal model capturing a Dirac electronic dispersion, both the $\beta$ and $\gamma$ pockets in SrFe$_2$As$_2$ and BaFe$_2$As$_2$ can be simultaneously explained by just two parameters (in addition to $c^\ast$). The unusual lemon-shaped topology of Fermi surface (illustrated in Fig.~\ref{FS}b) causes the extremal cross-sectional areas to exhibit cusp-like dependences on the orientation of the magnetic induction ${\bf B}$, accounting for the reported upturn in the magnetic field angle-dependence of the quantum oscillation frequencies~\cite{sebastian1,analytis1}. Further predictions are made for the angle-dependences of the cyclotron effective masses and magnetic breakdown tunneling effects, which we propose to provide an opportunity for studying excitations of strongly interacting fermions close to a Dirac point.

We begin by considering a dispersion of the form 
\begin{equation}\label{dirac}
\varepsilon=\pm\hbar c^\ast|k|+2t\cos[ck_z/2]+\mu,
\end{equation}
where $c$ is the bilayer spacing in the body-centered tetragonal crystal structure and a non-zero chemical potential $\mu$ causes asymmetry in size between electron and hole pockets~\cite{sebastian1,analytis1}. We neglect the anisotropy in $c^\ast$ identified in ARPES measurements~\cite{richard1} and band structure calculations~\cite{analytis1}, so that $k=\sqrt{k^2_x+k^2_y}$ where $k_x$ and $k_y$ lie within the FeAs-planes~\cite{splittingnote}. When ${\bf B}$ is aligned parallel to the crystalline $c$-axis and $k_z$, the extremal cross-sectional areas and cyclotron effective masses are 
\begin{equation}\label{area0}
A_{k,0}^\pm=\frac{\pi}{\hbar^2c^{\ast 2}}(2t\pm|\mu|)^2~~{\rm and}~~
m^\pm_0c^{\ast 2}=2t\pm|\mu|
\end{equation}
respectively, where the superscripts `$+$' and `$-$' refer to two different pocket sizes.

To determine the extremal Fermi surface cross-sections for an arbitrary orientation of ${\bf B}$, we introduce auxiliary coordinates $(k^\prime_y, k^\prime_z)={\bf R}_\theta(k_y, k_z)$, where ${\bf R}_\theta$ is a rotation matrix and $\theta$ is the angle between ${\bf B}\| k_z^\prime$ and the crystalline $c$ axis. Under such a transformation, the extremal orbits lie in the $(k_x, k^\prime_y)$ plane, with loci 
\begin{equation}\label{kx}
k_x^\pm=\sqrt{\bigg(\frac{2t}{\hbar c^\ast}\cos\bigg[\frac{ck^\prime_y\sin\theta}{2}\bigg]-\frac{\varepsilon\mp\mu}{\hbar c^\ast}\bigg)^2-k^{\prime 2}_y\cos^2\theta},
\end{equation}
obtained by setting $k^\prime_z=0$ and substituting $\pm\mu$ in place of $\mu$ in Eqn (\ref{dirac}). The cross-sectional areas and cyclotron effective masses are then obtained by evaluating
\begin{equation}\label{area}
A_{k,\theta}^\pm=4\int_0^{k_{\rm L}}\Re[k_x^\pm]{\rm d}k^\prime_y~~{\rm and}~~
m^\pm_\theta=\frac{\hbar^2}{2\pi}\frac{\partial A_k^\pm}{\partial\varepsilon}
\end{equation}
at $\varepsilon=0$, where $2k_{\rm L}=2\cos^{-1}[\pm|\mu|/2t]/c$ is the length of the pocket along $k_z$.

\begin{figure}
\centering
\includegraphics*[width=0.48\textwidth,angle=0]{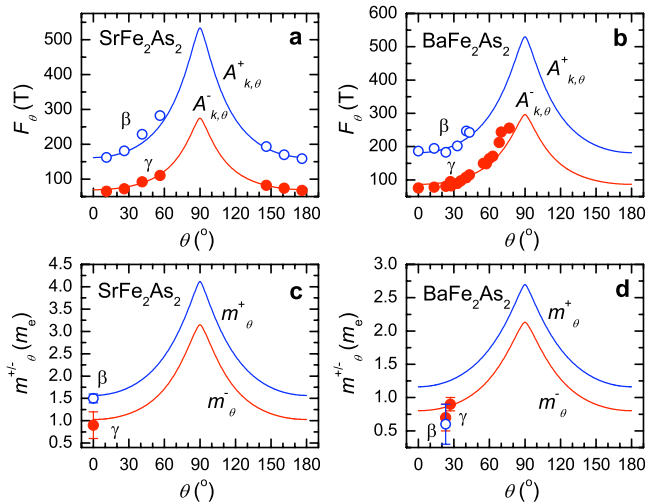}
\caption{Lines indicate simulated field angle-dependences of the magnetic quantum oscillation frequencies in SrFe$_2$As$_2$ ({\bf a}) and BaFe$_2$As$_2$ ({\bf b}) using Eqn.~(\ref{area}) and the Onsager relation, and the corresponding effective masses ({\bf c}) and ({\bf d}). The corresponding simulation parameters ($c^\ast=$~5.2~$\times$~10$^4$~ms$^{-1}$, $t=$~9.9~meV, $\mu=$~4.1~meV and $c=$~6.15~\AA~ for SrFe$_2$As$_2$ and $c^\ast=$~7.4~$\times$~10$^4$~ms$^{-1}$, $t=$~15.3~meV, $\mu=$~5.6~meV and $c=$~6.48~\AA~ for BaFe$_2$As$_2$) which determine the form of the dispersion given by Eqn.~\ref{dirac} are adjusted to match the experimental values (represented by circles) from Refs.~\cite{sebastian1,analytis1,Banote} and Fig.~\ref{MB}b.
}
\label{frequencies}
\end{figure}

We first turn to the topology of the $\beta$ and $\gamma$ pockets and make a comparison with vertically stacked electron and hole pockets expected for a Dirac-like dispersion. The experimentally measured angular dependence of the $\beta$ and $\gamma$ frequencies in both SrFe$_2$As$_2$ and BaFe$_2$As$_2$ is compared with predictions of the model in Eqn~1. The ratios $t/c^\ast$ and $\mu/c^\ast$ in the model can be optimised such that the $\theta$-dependences of the cross-sectional areas $A^\pm_{k,\theta}$ evaluated numerically (lines) using Eqn.~\ref{area} in Figs.~\ref{frequencies}a and b reproduce reasonably well the experimentally observed $\beta$ and $\gamma$ frequency angle dependences (circles) in both SrFe$_2$As$_2$ and BaFe$_2$As$_2$. The extremal cross-sectional areas are converted to quantum oscillation frequencies by use of the Onsager relation $F_\theta=(\hbar/2\pi e)A^\pm_{k,\theta}$. The ability of the model to reproduce simultaneously the experimental angular dependences of two frequencies with only two parameters (i.e. $t/c^\ast$ and $\mu/c^\ast$) represents a considerable improvement over the prior ellipsoidal approximation, which required four parameters in total (i.e. two parameters for each frequency~\cite{parameters}). The close agreement with experiment signals the $\beta$ and $\gamma$ pockets to be consistent with a single cone of Dirac fermions for which the Fermi energy intersects the Dirac point at a location in $k$ space. 

The next comparison we make is of the measured cyclotron effective masses with model predictions. The measured effective mass of the $\beta$-orbit (corresponding to $A^+_{k,\theta}$) in SrFe$_2$As$_2$ is larger than that of the $\gamma$-orbit (corresponding to $A^-_{k,\theta}$)~(see Fig.~\ref{frequencies}c,~\cite{Banote}). As anticipated for $\beta$ and $\gamma$ orbits arising from the same Dirac cone, the cyclotron effective mass is seen to be $A^\pm_{k,\theta}$-dependent (unlike for a quadratic dispersion)~$-$~(i.e.) higher for the larger cross-section $\beta$ orbit. A comparison of the measured cyclotron effective masses with simulations shown in Fig.~\ref{frequencies}c and d can yield individual estimates for all three parameters in the model (i.e. $t$, $\mu$ and $c^\ast$ in Eqn.~(\ref{dirac})). We use the effective mass data currently available at limited angles~\cite{sebastian1,analytis1} to obtain these estimates, presented in Figs.~\ref{frequencies}c and d. The estimate we obtain for $c^\ast$ (listed in Fig.~\ref{frequencies}) provides further support for the Dirac picture, proving consistent with ARPES experiments~\cite{richard1}.
%\begin{table}
%\begin{tabular}{ | l | l | l | l | l | l}
%        \hline
%        compound & $c^\ast$ (ms$^{-1}$) & $t$ (meV) & $\varepsilon_0$ (meV) & $c$ (\AA)\\ \hline\hline
%        SrFe$_2$As$_2$ & 5.2~$\times$~10$^4$ & 9.9 & 4.1 & 6.15\\ \hline
%        BaFe$_2$As$_2$ & 7.4~$\times$~10$^4$ & 15.3 & 5.6 & 6.48\\
%        \hline
% \end{tabular}\label{table}
% \caption{Parameters in Eqn.(\ref{dirac}) yielding consistency between the $\theta$-dependent Fermi surface cross-section areas simulated by Eqn.~(\ref{area}) and cyclotron effective masses simulated by Eqn.~(\ref{mass}) with experimental observations in both SrFe$_2$As$_2$ and BaFe$_2$As$_2$.}
% \end{table}

Given signatures of the validity of the Dirac distribution in antiferromagnetic SrFe$_2$As$_2$ and BaFe$_2$As$_2$ (Fig.~\ref{frequencies}), further magnetic quantum oscillation measurements are indicated to access the quasiparticle excitations close to the Dirac point. A test of a Dirac dispersion is typically provided by experimental verification of the effective mass vanishing as the quasiparticle trajectory approaches the Dirac point. In graphene, this is achieved by using electric fields to depopulate the pocket of carriers ~\cite{novoselov1}. The equivalent depopulation in SrFe$_2$As$_2$ and BaFe$_2$As$_2$ would require $\mu$ to be tuned by doping or pressure.

\begin{figure}
\centering
\includegraphics*[width=0.48\textwidth,angle=0]{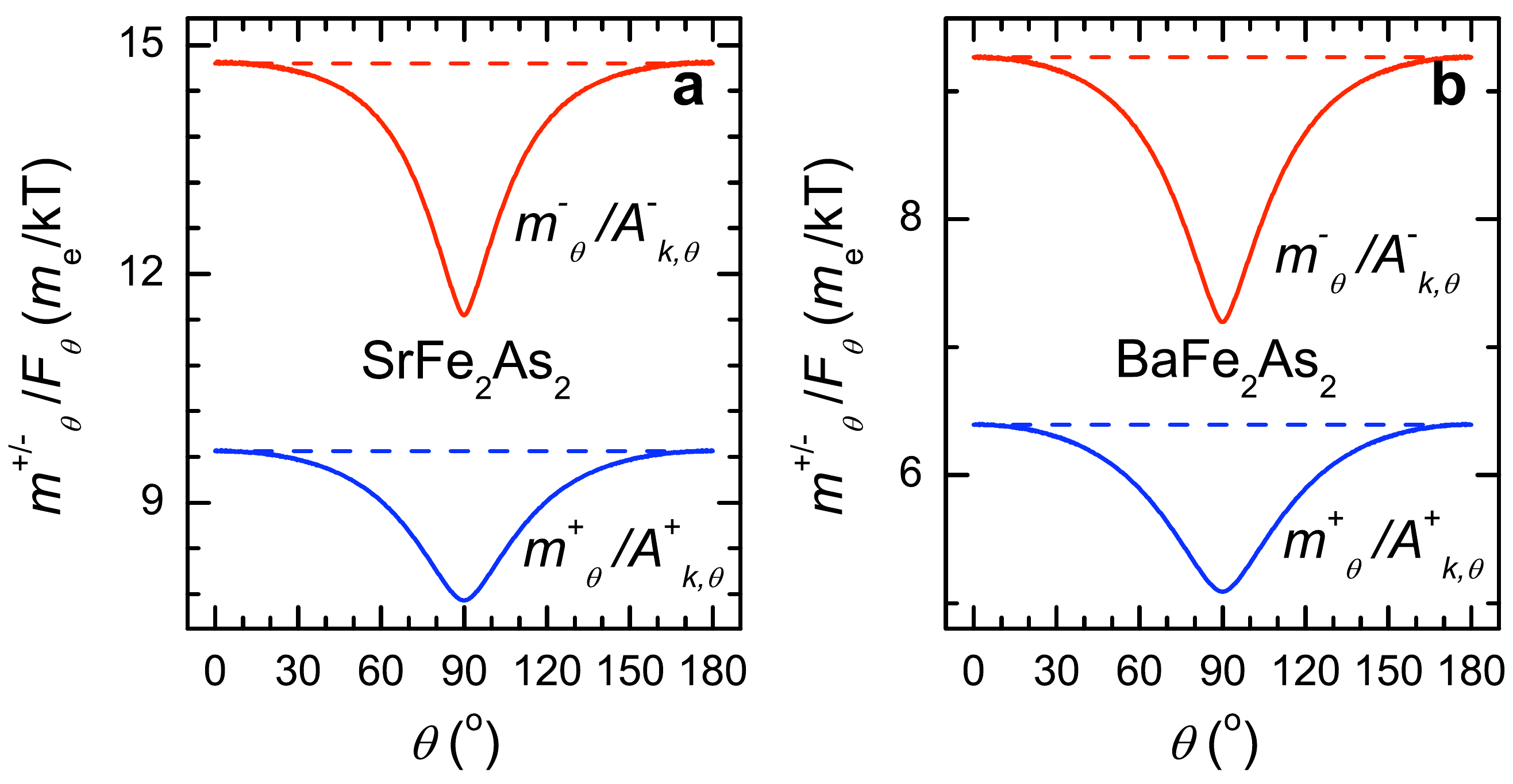}
\caption{Simulated ratio (solid lines) of the cyclotron effective mass $m^\pm_\theta$ to the frequency (given by the cross-sectional area $A^\pm_{k,\theta}$) for SrFe$_2$As$_2$ ({\bf a}) and BaFe$_2$As$_2$ ({\bf b}), yielding a pronounced dip at $\theta=$~90$^\circ$. The dip is a consequence of the vanishing contribution to the cyclotron mass from the extremity of the orbit closest to the Dirac point in $k$-space. For comparison, we show the same ratio expected for an ellipsoidal Fermi surface comprising conventional Landau quasiparticles (dashed lines), which is independent of $\theta$.
}
\label{ratio}
\end{figure}

Angle-dependent measurements provide an alternative means for accessing the Dirac point that does not require doping. As the angle $\theta$ is increased, the radius of curvature $k_{\rm R}$ (i.e. the radius of a circle tangent to the orbit and having the same curvature) at the furthest extremity of the orbit (shown schematically in Fig.~\ref{FS}b) undergoes a dramatic reduction in size, collapsing to zero as $\theta\rightarrow$~90$^\circ$. The high extremal quasiparticle velocity $v_{\rm E}=c^{\ast}\cos\theta+(tc/\hbar)\sin\theta\sin[(ck^\prime_y/2)\sin\theta]$ turning on a small radius translates to a vanishing contribution of a quantity we term the `extremal mass' $m_{\rm E}=\hbar k_{\rm R}/v_{\rm E}$ to the cyclotron mass at the farthest extremity of the orbit. Although $m_{\rm E}$ cannot be isolated in quantum oscillation experiments made at a single angle, the orbitally averaged cyclotron mass given by Eqn.~\ref{area} is strongly affected by $m_{\rm E}$. A consequence of $m_{\rm E}$ vanishing as $\theta\rightarrow$~90$^\circ$ is a marked reduction in the ratio $m^\pm_{\theta}/A^\pm_k$ near $\theta\sim 90^\circ$ in Fig.~\ref{ratio}. The predicted dip in this ratio near $\theta=$~90$^\circ$ (solid lines in Fig.~\ref{ratio}) is in contrast to the constant ratio $m_{\theta}/A=m_0/A_0$ expected for an ellipsoidal pocket of revolution comprising conventional Landau quasiparticles (dashed lines). Further experiments accessing the cyclotron mass as a function of angle are anticipated to provide a confirmation of the experimental distinction between massless Dirac quasiparticles in the antiferromagnetic FeAs family and conventional Landau quasiparticles.

\begin{figure}
\centering
\includegraphics*[width=0.48\textwidth,angle=0]{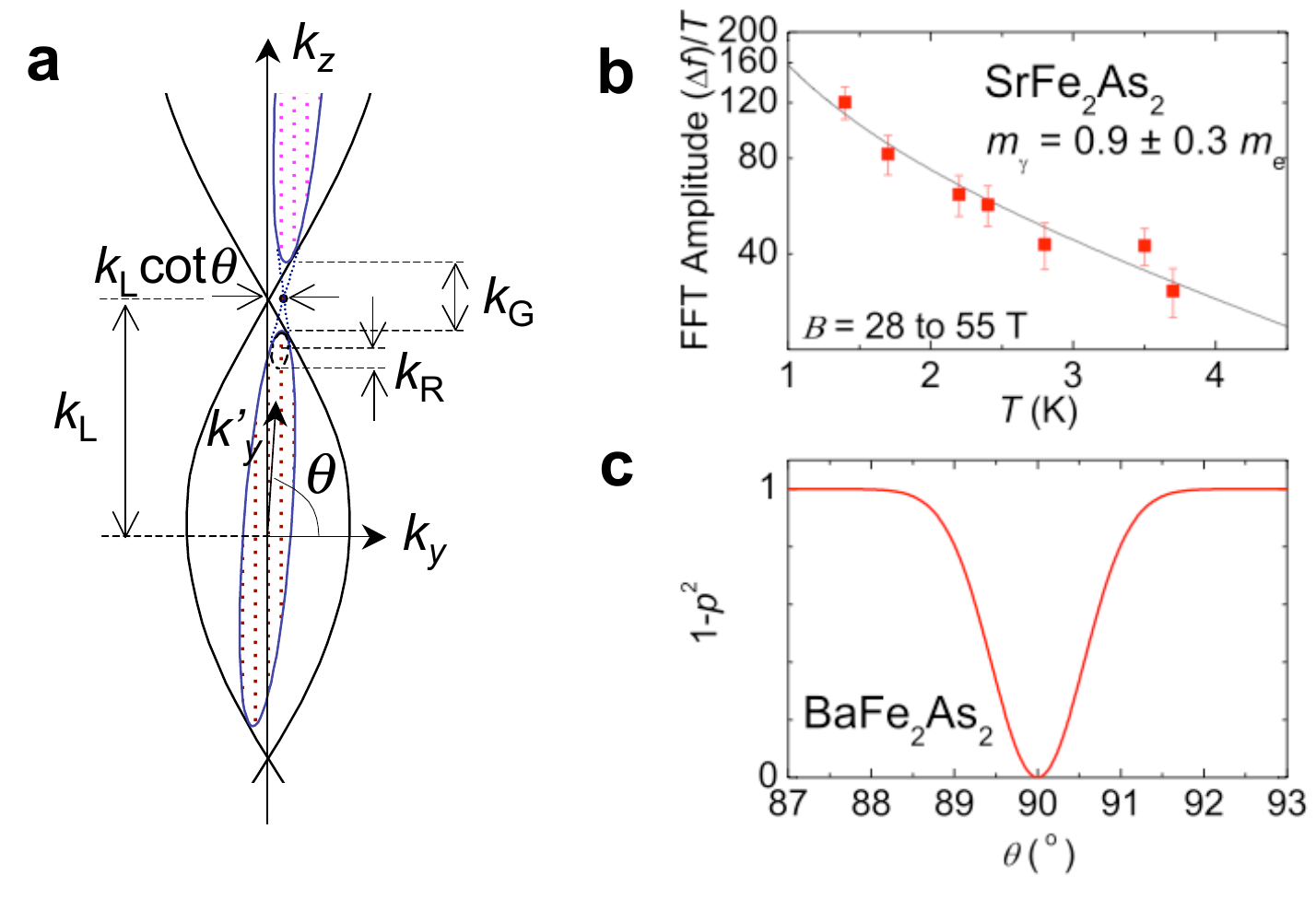}
\caption{{\bf a}, Schematic of the local geometry in the vicinity of.the Dirac point where magnetic breakdown occurs as $\theta\rightarrow$~90$^\circ$. {\bf b}, Amplitude of the $\gamma$ frequency in SrFe$_2$As$_2$ from Ref.~\cite{sebastian1} after Fourier transforming the raw data having subtracted a third order polynomial, with a fit to $aT/\sinh(14.69 m_\gamma T/B)$ to obtain the cyclotron mass $m_\gamma$. {\bf c} Amplitude attenuation factor $1-p^2$ calculated for the $\gamma$ pocket observed in BaFe$_2$As$_2$~\cite{analytis1}.
}
\label{MB}
\end{figure}

Finally, we note that once $\theta=$~90$^\circ$, the Fermi surface topology no longer supports closed extremal orbits. The quasiparticles instead move continuously between the electron and hole pockets via the Dirac point. Magnetic breakdown tunneling further ensures that open orbits will occur for a finite range of angles near 90$^\circ$. The probability $p^2=\exp(-B_0/B)$ of magnetic breakdown (across the gap shown schematically in Fig.~\ref{MB}a) depends on the characteristic magnetic breakdown field $B_0=(\pi\hbar/e)\sqrt{k^3_{\rm G}k_{\rm R}/2}$~\cite{shoenberg1} for this Fermi surface topology. This in turn depends on the band gap $k_{\rm G}$ at the point where magnetic breakdown breakdown occurs (i.e. the separation between electron and hole orbits for a given value of $\theta$, see Fig.~\ref{MB}a) and on the radius of curvature of the orbit $k_{\rm R}$$-$ both of which depend on the form of the dispersion. The dispersion in the immediate vicinity of the Dirac point for small $k_z-k_{\rm L}$ has the approximate form $\varepsilon\approx\pm\hbar c^\ast|k|+\hbar v_{\rm E}(k_z-k_{\rm L})$, from which we estimate $k_{\rm G}\approx2(c^\ast/v_{\rm E})k_{\rm L}\cot\theta$ and $k_{\rm R}\approx(v_{\rm E}/c^\ast)k_{\rm L}\cot\theta$ for a point of magnetic breakdown located at a distance $k_y=k_{\rm L}\cot\theta$ away from the $z$-axis in Fig.~\ref{MB}a. Substituting $k_{\rm G}$ and $k_{\rm R}$ into the above expression for $B_0$, we obtain
\begin{equation}\label{B0}
B_0\approx\frac{2\pi\hbar}{e}\bigg(\frac{c^\ast}{v_{\rm E}}\bigg)k^2_{\rm L}\cot^2\theta.
\end{equation}
Such a strongly angle-dependent magnetic breakdown field, which vanishes as $\theta\rightarrow$~90$^\circ$, is a unique feature of the Fermi surface topology in the vicinity of a Dirac point. While open orbits do not contribute to the magnetic quantum oscillations, there will be a loss of amplitude for the closed orbit given by $1-p^2=1-\exp(B_0/B)$. The anticipated form of $1-p^2$ for the case of the $\gamma$ pocket of BaFe$_2$As$_2$ is shown in Fig.~\ref{MB}c. The dramatic attenuation of the quantum oscillation amplitude of the $\gamma$ pocket expected near $\theta=$~90$^\circ$ in the model may account for its reported loss in Ref.~\cite{analytis1} at large angle.

In conclusion, we have shown from the angle-dependence of the magnetic quantum oscillations in SrFe$_2$As$_2$ and BaFe$_2$As$_2$ that the measured small Fermi surface pockets are consistent with Dirac nodal pockets arising from a Dirac dispersion with interlayer warping. We propose further angle-dependent experiments that can be performed to probe the massless nature of the quasiparticles approaching the Dirac point, and find that the quantum oscillations ultimately become attenuated at $\theta=$~90$^\circ$ due to magnetic breakdown tunneling giving rise to open orbits.

Our finding of Dirac fermions in SrFe$_2$As$_2$ and BaFe$_2$As$_2$ implies that the elementary excitations and orbital quantization effects are likely to be different from those in conventional spin-density wave materials$-$ becoming in fact more like those in graphene~\cite{neto1}. An intruiging aspect of the FeAs materials is not only the preservation of a metallic density of states despite a large spin density wave gap, but in fact the creation of Dirac nodes as a consequence of spin-density wave folding, thereby defeating complete Fermi surface gapping. The multi-orbital character of the FeAs family appears crucial in this creation of Dirac nodes at the meeting point of electron and hole dispersions of different orbital symmetry~\cite{ran1}. The virtual immunity of the observed pockets of Dirac fermions to localization effects could potentially explain the persistence of metallic behavior despite the very small density of carriers and strong correlations in the undoped antiferromagnetic materials. Of additional special interest is the greatly reduced relativistic speed $c^\ast\approx$~5$-$7~$\times$~10$^{4}$~ms$^{-1}$ in these materials compared to 10$^6$~ms$^{-1}$ in graphene, signalling the significantly stronger correlations of the Dirac fermions in SrFe$_2$As$_2$ and BaFe$_2$As$_2$. A pertinent question concerns whether strongly interacting Dirac fermions can tunnel through barriers or support superconducting pairs over long distances (by way of the proximity effect) in a similar manner to those in graphene~\cite{neto1}.

This work is supported by the US Dept. of Energy, the National Science Foundation and the State of Florida through the National High Magnetic Field Lab, Trinity College (U. of Cambridge), and the Royal Society. We acknowledge helpful discussions with M. Johannes.


\begin{thebibliography}{99}
\bibitem{Banote} The $\beta$ pocket in BaFe$_2$As$_2$~\cite{analytis1} is reported to have an anomalously low mass and large error bar, perhaps reflecting its small amplitude in Ref.~\cite{analytis1}.
\end{thebibliography}
\end{document}